\documentclass[useAMS,usenatbib]{mn2e}
\usepackage{graphicx}

\newcommand{\xmmn}{{\it XMM-Newton~\/}}
\newcommand{\asca}{{\it ASCA~\/}}

\newcommand{\suzaku}{{\it Suzaku~\/}}

\def\H0{{\rm ~km~s^{-1}~Mpc^{-1}}}

\def\la{\mathrel{\hbox{\rlap{\hbox{\lower4pt\hbox{$\sim$}}}{\raise2pt\hbox{$<$}}}}}
\def\ga{\mathrel{\hbox{\rlap{\hbox{\lower4pt\hbox{$\sim$}}}{\raise2pt\hbox{$>$}}}}}
\def\d25{D$_{25}$}

\def\deg{\hbox{$^\circ$~\/}}

\title[Suzaku observations of Markarian 335] {Suzaku observations of
  Markarian 335: evidence for a distributed reflector}

\author[J.~Larsson, et al.]  {\parbox[]{6.0in}
  {J. Larsson~$^1$\thanks{E-mail: jlarsson@ast.cam.ac.uk},
    G. Miniutti~$^1$, A. C. Fabian~$^1$, J. M. Miller~$^2$,\\
    C. S. Reynolds~$^3$ and G. Ponti~$^{4,5}$\\
    \footnotesize {\it $^1$Institute of Astronomy, University of
      Cambridge, Madingley
      Road, Cambridge CB3 0HA\\
      $^2$Department of Astronomy and Astrophysics, The University of
      Michigan,
      500 Church Street, Ann Arbor, Michigan, 48109, USA\\
      $^3$Department of Astronomy, University of Maryland, College
      Park,
      MD 20742, USA\\
      $^4$Dipartimento di Astronomia dell' Universit\'{a} degli Studi
      di Bologna,
      via Ranzani I, I--40127, Bologna, Italy \\
      $^5$IASF/INAF Sezione di Bologna, via Gobetti 101, I--40129,
      Bologna, Italy\\}}}

\date{Accepted 2007 December 4. Received 2007 November 15; in original
  form 2007 October 16}

\pagerange{\pageref{firstpage}--\pageref{lastpage}}
\pubyear{2007}

\begin{document}

\maketitle

\label{firstpage}

\begin{abstract}
  We report on a 151~ks net exposure \suzaku observation of the Narrow
  Line Seyfert 1 galaxy Mrk~335. The 0.5--40~keV spectrum contains a
  broad Fe line, a strong soft excess below about 2~keV and a Compton
  hump around 20--30~keV. We find that a model consisting of a power
  law and two reflectors provides the best fit to the time-averaged
  spectrum. In this model, an ionized, heavily blurred, inner
  reflector produces most of the soft excess, while an almost neutral
  outer reflector (outside $\sim 40\ r_g$) produces most of the Fe
  line emission. The spectral variability of the observation is
  characterised by spectral hardening at very low count rates. In
  terms of our power-law + two-reflector model it seems like this
  hardening is mainly caused by pivoting of the power law. The rms
  spectrum of the entire observation has the curved shape commonly
  observed in AGN, although the shape is significantly flatter when an
  interval which does not contain any deep dip in the lightcurve is
  considered. We also examine a previous 133~ks \xmmn observation of
  Mrk~335. We find that the \xmmn spectrum can be fitted with a
  similar two-reflector model as the \suzaku data and we confirm that
  the rms spectrum of the observation is flat. The flat rms spectra,
  as well as the high-energy data from the \suzaku PIN detector,
  disfavour an absorption origin for the soft excess in Mrk~335.

 \end{abstract}


\section{Introduction}

The X-ray spectra of type 1 Active Galactic Nuclei (AGN) are usually
dominated by a power-law component. They also often exhibit a broad,
skewed Fe line around 6 keV, as well as smooth component rising above
the power-law continuum below about 2~keV, the so called soft
excess. While it is well established that the broad Fe line arises due
to reflection of the continuum power law in the inner accretion disc
(where it is broadened by Doppler and relativistic effects), the
origin of the soft excess remains a question of much debate.

The soft excess has a very similar spectral shape in AGN covering
several decades in mass (Gierli\'{n}ski \& Done 2004; Crummy et
al. 2006), suggesting that its origin is in atomic rather than thermal
processes. An obvious candidate of atomic origin is the reflection
spectrum associated with the broad Fe line. In this model, the soft
excess is due to the blurring of soft emission lines from the inner
parts of the accretion disc.  The reflection model has been applied
successfully in a number of AGN (e.g. Crummy et al. 2006), although
sources that exhibit a strong soft excess but no broad Fe line are
difficult to explain in this model (see e.g. Brenneman et
al. 2007). Another proposed explanation for the soft excess is that it
is due to smeared absorption by optically thin material in the line of
sight (Gierli\'{n}ski \& Done 2004).  This model has been seen to
provide equally good fits as the reflection model (e.g. Middleton et
al. 2007). However, the model used in these fits has been very
simplistic, ignoring emission resulting from the absorber and the
acceleration of the matter.  A more realistic absorption model does
not reproduce the smooth shape of observed soft excesses unless
extreme velocities are incorporated (Schurch \& Done 2007).

Mrk 335, also known as PG 0003+199, is a Narrow Line Seyfert 1 (NLS1)
galaxy at redshift $z=0.026$. It exhibits both a strong soft excess
and a broad Fe line. Unlike most Seyfert 1 galaxies, it does not show
any clear signs of complex, warm absorption at low energies, making it
an ideal target for studying the origin of the soft excess.

The soft excess in Mrk~335 was first observed by {\it EXOSAT} (Pounds
et al. 1986; Turner \& Pounds 1988) and was later confirmed by BBXRT
(Turner et al. 1993).  No significant absorption edges were detected
in the BBXRT observation but an Fe edge in {\it Ginga} data lead
Turner et al. (1993) to suggest the presence of a variable, ionized
absorber. A subsequent \asca observation again revealed the presence
of a strong soft excess, but showed no clear evidence of edges or
ionized absorption at low energies (Reynolds 1997; George et
al. 1998). The \asca spectrum also showed evidence for a broad Fe line
(Nandra et al. 1997), and Ballantyne et al. (2001) found a good fit to
the entire 0.6--10~keV band with an ionized reflection model.  {\it
  BeppoSAX} data (Bianchi et al. 2001) clearly showed a broad Fe line
as well as a soft excess, but revealed a rather small Compton
reflection component.

\xmmn observed Mrk~335 for the first time in 2000. The observation was
first analysed by Gondoin et al. (2002), who favour a model in which
the soft excess consists of a combination of Bremsstrahlung emission
and ionized reflection from the accretion disc. No absorption edges
apart from the 0.54~keV edge due to Galactic Oxygen were seen in the
RGS spectrometer.

The \xmmn observation was later re-analysed by Longinotti et
al. (2007), who detected a narrow absorption feature at 5.9~keV,
which, if it is identified with Fe {\scriptsize XXVI}, is inflowing at
$\sim 0.11-0.15\ c$, or, if at rest, is located close to the black
hole and gravitationally redshifted. Both Crummy et al. (2006) and
Middleton et al. (2007) have also fitted the \xmmn data as a part of a
sample of type 1 AGN. Crummy et al. (2006) find that the spectrum can
be well fitted with a reflection model, while Middleton et al. (2007)
show that an equally good fit can be obtained if the soft excess is
modelled as smeared absorption.

In January 2006, \xmmn re-observed Mrk~335 for 133~ks. In addition to
a strong soft excess below 2~keV, this observation revealed a
double-peaked Fe emission feature with peaks at 6.4 and 7~keV (O'Neill
et al. 2007). Spectral fitting of this feature suggested that a
moderately broad relativistic line is present but that the two peaks
are due to narrow lines originating in more distant material. The
authors further note that the reflection model inferred from the data
above 3~keV still leaves a soft excess when extrapolated to lower
energies. The rms variability spectrum of the observation is
consistent with being constant, disfavouring the absorption model for
the soft excess, which, assuming that the ionization state of the
absorber is driven by the continuum variability, predicts enhanced
variability at low energies (Gierli\'{n}ski \& Done 2006).

Recent {\it Swift} observations of Mrk~335, carried out in May and
June/July 2007, caught the source in an extremely low state, with the
flux having diminished by a factor of 30 compared to previous
observations. In this low state the source exhibited a very hard
spectrum above 2~keV as well as a soft excess at low energies. The
spectral changes together with the sudden drop in flux have been
interpreted in terms of a partial covering model (Grupe et al. 2007)
as well as in a reflection model (Gallo et al. 2007, in prep).

In this paper we present the results of a 316~ks ($\sim 150$~ks net
exposure) \suzaku observation of Mrk~335 performed in June 2006. This
observation was carried out before the sudden drop in flux mentioned
above and caught the source in its typical flux state. We also analyse
the 133~ks \xmmn observation performed in January 2006.

This paper is organised as follows: Section \ref{observations}
describes the observations and the data reduction, Sections
\ref{timeav} and \ref{xmmspec} describe the spectral analysis of the
\suzaku and \xmmn time-averaged spectra, and the spectral variability
of both observations is presented in Section \ref{variability}.
Section \ref{discussion} contains a discussion of the results, and a
summary is given in Section \ref{summary}.

\section{Observations and data reduction}
\label{observations}

\subsection{The Suzaku observation}

Mrk 335 was observed by \suzaku between 2006 June 21--24 for a total
duration of 316 ks. Event files from version 1.2.2.3 of the {\it
  Suzaku} pipeline processing were used and spectra were extracted
using {\scriptsize XSELECT}.

The net exposure time of all four XIS detectors is 151~ks. For each
XIS, source spectra were extracted from circular regions of 4.3 arcmin
radius centred on the source (which was observed off-axis in the HXD
nominal position). Background spectra were extracted from two circular
regions with the same total area as the source region, avoiding the
chip corners with the calibration sources. Response matrices and
ancillary response files were generated for each XIS using
{\scriptsize XISRMFGEN } and {\scriptsize XISSIMARFGEN} version
2006-11-26. The ARF generator should account for the hydrocarbon
contamination on the optical blocking filter (Ishisaki et al. 2007).

The source count rates over the 0.7--10~keV energy range for the three
front-illuminated (FI) detectors are $1.299\pm 0.003\ \rm{counts\
  s}^{-1}$(XIS0), $1.484\pm 0.003\ \rm{counts\ s}^{-1}$ (XIS2) and
$1.364\pm 0.003\ \rm{counts\ s}^{-1}$ (XIS3). The back-illuminated
(BI) detector (XIS1) has a count rate of $2.243\pm 0.004\ \rm{counts\
  s}^{-1}$ in the 0.3--8~keV band. The background typically
contributes 2--3 per cent of the total count rate for all four
detectors.

At the time of our observation, the bias voltage for 16 of the 64 HXD
PIN diodes had been reduced from 500 V to 400 V, in order to suppress
a rapid increase of noise events. The PIN event file was therefore
filtered to exclude the diodes biased with 400 V, and the
corresponding response file (ae\_hxd\_pinhxnom123\_20060814.rsp) was
used.  A model for the non-X-ray background was provided by the HXD
team. Source and background spectra were constructed from identical good
time intervals and the exposure time of the background spectrum was
increased by a factor of 10 (to account for the fact that the
background model was generated with 10 times the actual count rate in
order to minimise the photon noise). After deadtime correction the net
exposure time of the PIN was 120~ks.

The total PIN count rate over the 14--40 keV energy range is $0.296
\pm 0.002\ \rm{counts\ s}^{-1}$, compared to $0.262 \pm 0.005\
\rm{counts\ s}^{-1}$ for the background. Since the background is so
much higher than the source, the accuracy of the background model is
very important for our results.  We therefore compared the background
model to the night earth spectrum, which was obtained by selecting
periods of earth occultation during the observation. Since the earth
is dark in the hard X-rays the background model should agree with the
night earth spectrum. We find that that the background model
overpredicts the night earth data by about 6 per cent, but that the
spectral shape is well reproduced. We can therefore simply correct the
count rate of the background model so that it matches the night earth
data when fitting the time-averaged spectrum.  Investigations of the
background model on shorter time-scales unfortunately show that the
systematic uncertainties of the model are too large for timing
analysis to be carried out, and we therefore defer this analysis to a
later time when a more accurate background model will be available.

Since the background model discussed above does not include the
contribution from the cosmic X-ray background (CXB), a spectral model
for it is included in all the fits. The model, which has the form
$2.06 \times 10^{-6} (E/100\ \rm{keV})^{-1.29}\ exp{(-E/41.13\
  \rm{keV})}$, is based on the HEAO-A1 spectrum and has been
renormalized to the HXD field of view. The CXB flux over the 14--40
keV band predicted by this model represents about 25 per cent of the
source flux.

\subsection{The XMM-Newton observation}

\xmmn observed Mrk~335 between 2006 January 03--05 for a total
duration of 133~ks (observation ID 0306870101). The data were reduced
using the \xmmn Science Analysis System version (7.0.0). We use only
the data from the EPIC pn camera, which was operated in the Small
Window mode during the observation.  Because of the Small Window mode
the data are not affected by photon pile-up.  The source spectrum was
extracted from a circular region of radius 34 arcsec centred on the
source, and the background spectrum was extracted from two circular
regions with a total area three times that of the source region.
Inspection of the background showed that strong flaring was present at
the beginning and at the end of the observation.  Excluding these time
intervals leaves a total good exposure time of 115~ks.

\section{The time-averaged Suzaku spectrum}
\label{timeav}

In order to investigate the agreement between the four XIS detectors
we initially fitted them separately with a simple power-law model.
The three FI XIS showed very similar spectra over the 0.7--10 keV
energy range, with the photon indices agreeing to within 0.01.  Larger
differences are seen below 0.7~keV and the agreement becomes very poor
below about 0.6~keV (see e.g. Fig. \ref{se_ratio}). Since we find that
including the data between 0.6 and 0.7~keV only makes the quality of
the fits worse without significantly changing any of the fit
parameters, we choose 0.7~keV as our lower energy limit. (We will use
the BI XIS to constrain our spectral models at lower energies.) When
the 0.7--10~keV FI XIS spectra were fitted simultaneously with the
photon index tied, the relative normalizations were found to be
XIS0/XIS2~$=0.93$ and XIS3/XIS2~$=0.99$. These values were also
obtained when fitting more complicated models, and in all subsequent
fits we will fit the FI XIS simultaneously with gamma tied. The BI
XIS (XIS1) was found to have a somewhat steeper spectrum than the
other XIS ($\Delta \Gamma \approx 0.1$ in the same energy range). As
XIS1 also has a higher effective area at softer energies (we use the
0.3--8 keV energy range) we choose to fit it separately and use it
mainly to constrain our models at very low energies. Due to
calibration uncertainties near the instrumental Si K edge at 1.84 keV
we always ignore 1.8--1.9~keV in the FI XIS and 1.7--1.9~keV in the
BI XIS.

The PIN spectrum is unfortunately associated with uncertainties
regarding the background modelling and the cross normalization with
respect to the XIS detectors. Although this means that we cannot use
the PIN data to accurately constrain parameters of different spectral
models, we can still use it to distinguish between models that fit the
data at lower energies.

Below we will first consider the 2--10~keV spectrum and then include
the low-energy data down to 0.3~keV. As a last step we will consider
the 14--40~keV PIN spectrum. Throughout this paper, errors are quoted
at the 90 per cent confidence level and energies of spectral features
are quoted for the rest frame of the source.

\subsection{The 2 -- 10 keV  spectrum}
\label{2to10}

\begin{center}
\begin{figure}
\rotatebox{270}{\resizebox{!}{80mm}{\includegraphics{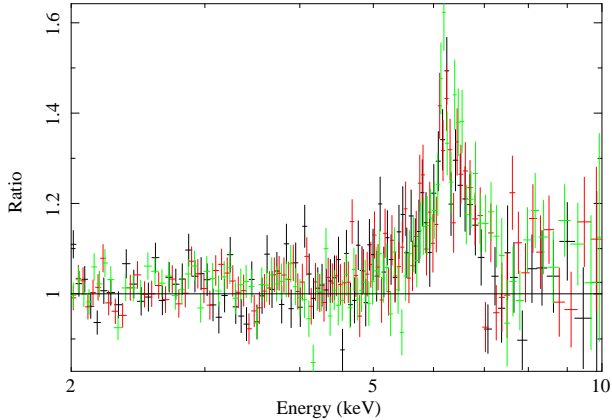}}}
\caption{The \suzaku FI XIS spectrum shown as a ratio to a power law
  fitted over the 2--4.5 and 7.5--10 keV energy ranges. Data points in
  black, red and green correspond to XIS 0, 2 and 3.}\label{fe_ratio}
\end{figure}
\end{center}

The most prominent feature in the 2--10~keV energy range is the broad
Fe line which extends from about 5 to 7~keV. The line profile is shown
in Fig.  \ref{fe_ratio} as a ratio to a power law modified by Galactic
absorption ($\rm {N_H = 3.99 \times 10^{20}\ cm^{-2}}$, included in
all fits from hereon), fitted over the 2--4.5 and 7.5--10 keV energy
ranges. The power law has a photon index of $\Gamma = 2.09 \pm 0.02$.

In order to find a phenomenological description for the data over the
whole 2--10~keV band we start by adding a Gaussian line to the
power-law model. This simple model provides a good fit to the data
with $\chi^2=4441$ for 4353 degrees of freedom (d.o.f.). The Gaussian
line is found to have $\rm{E=6.43 \pm 0.04\ keV}$, $\sigma =
0.40^{+0.05}_{-0.04}\ \rm{keV}$ and equivalent width $\rm{EW = 285\pm
  34\ eV}$. The photon index of the power law is still $\Gamma = 2.09
\pm 0.01$. When adding a narrow line around 6.4~keV to the model (to
check for a narrow component of the Fe line from distant matter) the
fit improves by $\Delta \chi^2 =19$ for 2 degrees of freedom.  The
narrow line has energy $\rm{E=6.36\pm 0.03\ keV}$\footnote{After
  completion of this work a new processing version of these data was
  released (V.2.0.6.13). With this version the energy of the narrow
  line is found to be $\rm{E=6.39^{+0.02}_{-0.03}\ keV}$, i.e
  completely consistent with the 6.4~keV Fe line. No other model
  parameters changed significantly with the new processing version.}
and equivalent width $\rm{EW =29\pm 14\ eV}$. With the narrow line
included, the broad line parameters change to $\rm{E=6.45 \pm 0.05\
  keV}$, $\sigma = 0.47^{+0.10}_{-0.06}\ \rm{keV}$ and $\rm{EW =
  250^{+40}_{-39}\ eV}$.

No other narrow lines are detected in the spectrum. In particular, we
do not detect the 5.9~keV absorption line seen in the 2000 \xmmn
observation (Longinotti et al.  2007) or the 7~keV emission line seen
in the 2006 \xmmn observation (O'Neill et al. 2007). If the narrow
line around 7~keV originates in distant material, as suggested by
O'Neill et al. (2007), it seems strange that is has disappeared in the
6 months between the \xmmn and \suzaku observations. However, adding a
line around 7~keV to our model for the \suzaku data does not improve
the quality of the fit, and we find that the line tends to move to
energies much lower than 7~keV. If we freeze the line energy at
6.97~keV (the energy of Fe {\scriptsize XXVI} Ly$\alpha$) we find a
line flux of $0.83^{+1.8}_{-0.8}\times 10^{-6} \rm{photons\ cm^{-2}\
  s^{-1}}$, corresponding to an equivalent width of
$\rm{EW}=6^{+15}_{-6}$~eV. Using the same model for the 2006 \xmmn
data, we find a significantly higher line flux of
$4.96^{+1.40}_{-1.82}\times 10^{-6} \rm{photons\ cm^{-2}\ s^{-1}}$. It
thus seems clear that the emission line around 7~keV has disappeared
or at least weakened significantly in the \suzaku observation.

We now move on to try and find a more realistic model for the data.
The broad, skewed Fe line seen in Fig.  \ref{fe_ratio} arises as the
incident power law is reflected in the accretion disc, and should
therefore have a reflection continuum associated with it. To model the
reflection from an accretion disc we use the constant-density model
{\scriptsize REFLION} by Ross \& Fabian (2005).  The parameters of the
model are the Fe abundance, the ionization parameter $\xi$, the photon
index of the incident power law and the normalization. In order to
account for relativistic effects in the vicinity of the black hole,
this model is convolved with the relativistic blurring kernel
{\scriptsize KDBLUR}, which is derived from the code by Laor (1991).
The relativistic blurring parameters are the inner and outer radii of
the disc ($r_{\rm in}$ and $r_{\rm out}$), the inclination $i$ and the
emissivity index $q$ (the emissivity follows the form $\epsilon
\propto r^{-q}$, where $r$ is the radius of emission).

In addition to the power law and the blurred reflection component, we
also include the narrow Fe K$\alpha$ line in our model. In all fits we
fix the Fe abundance at the solar value, the emissivity index at $q=3$
and the outer radius of the disc at $r_{\rm out} = 400\ r_g$ (where
$r_g=GM/c^2$ is the gravitational radius).  We find a best fit of
$\chi^2/\rm{d.o.f.} = 4416/4353$ with $\Gamma=2.19\pm 0.01$,
$\xi=30^{+3}\ \rm{erg\ cm\ s^{-1}}$ (the lowest allowed value),
$i=54\deg^{+8}_{-6}$ and $r_{\rm{in}}=40^{+60}_{-20}\ r_g$. The
parameters of the narrow line are found to be $\rm{E=6.36\pm 0.02\
  keV}$ and $\rm{EW =60^{+14}_{-15}\ eV}$. In order to confirm these
results we also fitted a model in which we replaced the blurring
kernel {\scriptsize KDBLUR} with the {\scriptsize KERRCONV} model by
Brenneman \& Reynolds (2006). An important difference between these
two models is that the {\scriptsize KERRCONV} model has the spin of
the black hole as a free parameter.  The {\scriptsize KERRCONV} model
gives best-fitting parameters that are consistent with those reported
above, and gives a best-fitting spin value of 0, as expected from the
relatively large value of $r_{\rm{in}}$.

It should be noted that the best fit presented above is not
necessarily unique. Both a highly ionized disc and a high
inclination makes the blue peak of the Fe line bluer, and these
parameters are hence somewhat degenerate. In fact, with an ionisation
parameter of $\xi=2000\ \rm{erg\ cm\ s^{-1}}$, we find a fit of
comparable quality to the one above ($\chi^2/\rm{d.o.f.} = 4424/4353$)
with $i=19\deg^{+10}_{-19}$, $\Gamma=2.06\pm 0.01$ and similar
parameters for the narrow line. We will hopefully be able to break
this degeneracy by including the low-energy data.

\subsection{The soft excess}\label{softex}
\begin{center}
\begin{figure}
\rotatebox{270}{\resizebox{!}{80mm}{\includegraphics{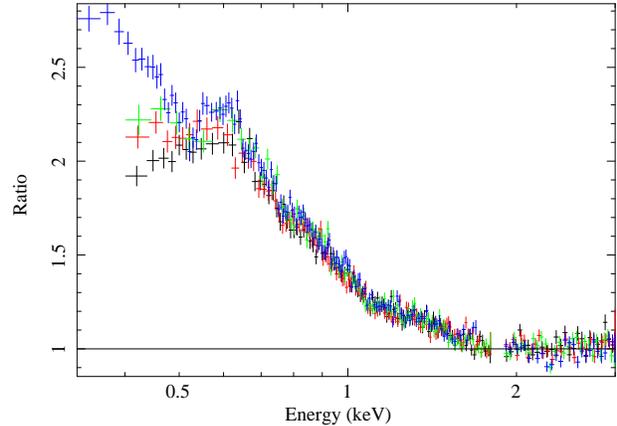}}}
\caption{The \suzaku XIS spectrum shown as a ratio to the 2--10~keV
  power-law + Gaussian model, clearly revealing the soft excess.  Data
  from the FI XIS are shown in black, red and green, and data from
  the BI XIS (XIS1) are shown in blue. The agreement between the
  detectors clearly becomes poor at low energies.  The structure seen
  in the 0.5-0.7~keV range in XIS1 is discussed in the
  text. }\label{se_ratio}
\end{figure}
\end{center}
\begin{figure}
\begin{center}
\rotatebox{270}{\resizebox{!}{80mm}{\includegraphics{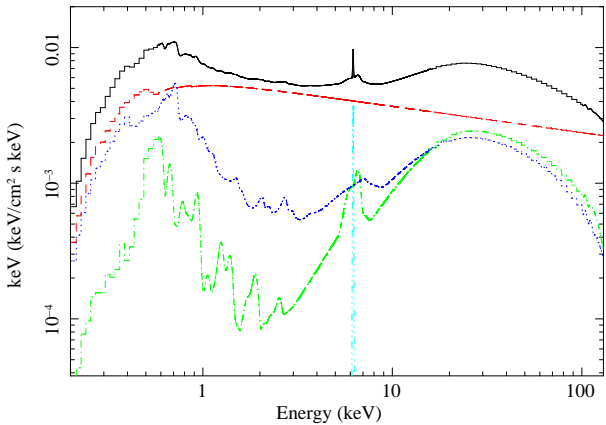}}}
\caption{Best-fitting two-reflector model for the time-averaged
  0.7--10 keV \suzaku spectrum. The solid, black line shows the
  total model and the dashed, red line shows the primary power law.
  The inner and outer reflectors are represented by the blue, dotted
  line and the green, dashed-dotted line, respectively. The model
  also includes a narrow Fe K$\alpha$ line.
}\label{2refmodel}
\end{center}
\end{figure}
\begin{figure}
\begin{center}
\rotatebox{270}{\resizebox{!}{80mm}{\includegraphics{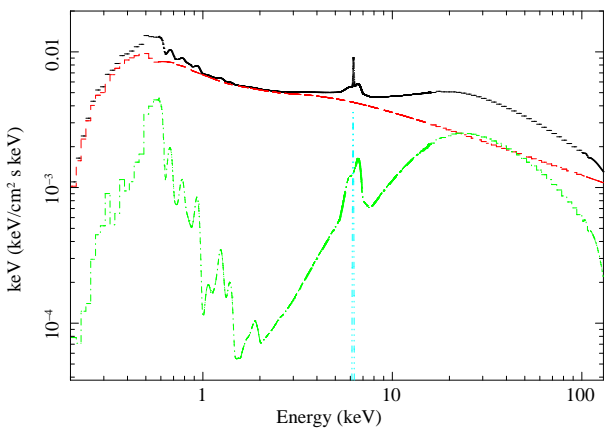}}}
\caption{ The best-fitting blurred absorption model for the
  time-averaged 0.7--10 keV \suzaku spectrum. The solid, black line
  shows the total model and the dashed, red line shows the absorbed
  power law. The model also includes distant reflection (green,
  dashed-dotted line) and a narrow Fe K$\alpha$ line.}\label{swmodel}
\end{center}
\end{figure}
We now include the data from the three FI XIS down to 0.7~keV (this
lower energy limit is discussed at the beginning of the section),
ignoring the 1.8--1.9 keV band which is affected by the instrumental
Si edge.  When extrapolating the best-fitting 2--10~keV models, a
clear, rather smooth soft excess is visible below 2~keV, as shown in
Fig.  \ref{se_ratio}. The ratio of the data to the model at 0.7~keV is
1.9 for the simple power-law model, and 1.7 and 1.5 for the highly
ionized and almost neutral reflection models, respectively.

A good fit to the soft excess can be found by simply adding a
black-body component to the 2--10~keV models.  A black-body of
temperature 0.13~keV together with our reflection model with $\xi=30\
\rm{erg\ cm\ s^{-1}}$ for instance gives $\chi^2/\rm{d.o.f.} =
5579/5340 $. Although this solution provides a good parametrisation of
the data it is very unlikely to be realistic. The temperature is
simply too hot for an accretion disc around a supermassive black hole,
and black-body temperatures of 0.1-0.2~keV have been found to fit soft
excesses in AGN with a large range of masses and accretion rates
(Gierli\'{n}ski \& Done 2004; Crummy et al. 2006), in disagreement
with the prediction that the disc temperature should scale as $T
\propto M^{-1/4}(\dot{M}/\dot{M}_{Edd})^{1/4}$ (Shakura \& Syunyaev
1973).

A natural explanation for the constant temperature of the soft excess
is that it is due to some atomic process, and reflection is thus an
obvious candidate. Since no emission lines are seen in the soft
excess, this reflection has to be heavily blurred, and therefore
originate in the very inner parts of the accretion disc. This poses a
problem for our 2--10~keV reflection models. The inner radius of
emission at $40\ r_g$, as suggested by the Fe line shape, simply does
not provide enough blurring to reproduce the smooth shape of the soft
excess. A disc truncated at $\sim 40\ r_g$ is also at odds with
Mrk~335 being a high accretion rate source
($\rm{\dot{M}/\dot{M}_{Edd}=0.79}$, based on Woo \& Urry 2002). It is
thus possible that the disc extends further in but is unable to
produce fluorescent Fe emission in the inner parts (e.g. because of
its ionization state).

It seems plausible that a model with two reflectors can reproduce the
data; an inner, heavily blurred reflector to produce the soft excess,
and an outer reflector to produce the sharp features of the Fe line.
We therefore construct a model with an inner reflector extending from
from 1.24 to $40\ r_g$ and an outer reflector extending from $40$ to
$400\ r_g$. If we fix the inclination at $i=54\deg$ (as found for 
our best fit to the 2-10~keV data) we find a fit with
$\chi^2/\rm{d.o.f.} = 5592/5335$. This model has a photon index of
$\Gamma=2.19\pm 0.01$ and inner and outer ionization parameters of
$\xi_{\rm in}=310^{+10}_{-8}\ \rm{erg\ cm\ s^{-1}}$ and $\xi_{\rm
  out}=32^{+5}_{-2}\ \rm{erg\ cm\ s^{-1}}$. The narrow line is found
to have energy $E=6.36\pm 0.02$~keV and an equivalent width of
$EW=67^{+16}_{-14}\ \rm{eV}$. The reflection fraction of the two
reflectors is 2.8, with each of the reflectors contributing roughly
the same amount. The model is plotted in Fig.  \ref{2refmodel}.

It is clear that this three-component model leaves much freedom for
many of the parameters and that the fit presented above is not
necessarily unique.  The fact that the inner disc needs to be more
highly ionized than the outer one however seems robust. We were for
example not able to find a satisfactory fit with a highly ionized
outer reflector (together with a lower inclination) as suggested by
one of our fits to the 2--10 keV data.

Another possible way of producing the soft excess is by
relativistically blurred absorption of a steep continuum. Just as for
the case of reflection, this model explains both the constant
temperature and the smooth shape of the soft excess.  In order to
investigate this possibility we use the {\scriptsize SWIND1} model of
Gierli\'{n}ski \& Done (2006). This model has three free parameters,
the velocity smearing $\sigma$ (assumed to be Gaussian), together with
the ionization parameter and the column density of the absorbing
material ($\xi_a$ and $N_H$ respectively). To model the continuum we
use a power law and a blurred reflection component, as the latter is
still required to explain the broad Fe line. For the reflection
component we use the same parameters as those found for the outer
reflector in the two-reflector model, leaving only the normalization
as a free parameter.  We find a fit of comparable quality as the
two-reflector model ($\chi^2/\rm{d.o.f.}  =5581/5334$) with
$\xi_a=499^{+53}_{-54}\ \rm{erg\ cm\ s^{-1}}$, $N_H=7.4\pm 0.9\times
10^{22}\ \rm{cm^{-2}}$, $\sigma=0.50_{-0.02}\ c$ (highest allowed
value) and $\Gamma=2.47^{+0.01}_{-0.02}$.  The model is shown in Fig.
\ref{swmodel}.

The fact that the soft excess can be equally well fitted with smeared
absorption as with blurred reflection from an accretion disc is
commonly observed in AGN (e.g. Middleton et al. 2007). However, the
velocity smearing of 0.5~$c$ that we find in our best fit with
the absorption model is unreasonably high. It should also be noted
that the {\scriptsize SWIND1} absorption model is rather simplistic in
its treatment of the velocity field (assumed to be Gaussian) and in
that it does not include any emission from the absorbing material. It
has been shown that when these issues are addressed, the model spectra
include sharp features which contrast strongly with the observed
smooth shapes of soft excesses (Schurch \& Done 2007). We therefore
favour the two-reflector model presented above as an explanation for
the soft excess in Mrk~335.

\subsubsection{The spectrum from the back-illuminated XIS1 detector}

For XIS1, we use the 0.3--8~keV energy band, ignoring the 1.7-1.9~keV
range due to the effects of instrumental Si. Since the calibration of
XIS1 is still somewhat uncertain we start by comparing its low-energy
spectrum with that from the long \xmmn observation. When taking the
ratio of the two data sets, clear, sharp differences can be seen at
low energies. The structure seen in the 0.5--0.7~keV range in
Fig. \ref{se_ratio} is for instance not present in the \xmmn data. As
it is quite likely that at least some of these differences are due to
calibration uncertainties in XIS1, we will only very briefly discuss
the XIS1 spectrum.

The two-reflector model provides a good fit to the XIS1 data down to
0.5~keV ($\chi^2/\rm{d.o.f.}=1614/1605$), with none of the parameters
changing significantly from the values presented above. Extrapolation
of the model down to 0.3~keV however reveals a soft excess as well as
the very sharp edge around 0.5~keV.  In order to fit these features
with the two-reflector model we need to add an edge at 0.49~keV, and
allow for a steeper emissivity profile ($q=6.5$) as well as a steeper
power law ($\Gamma=2.29$).  If the 0.49~keV edge is intrinsic to the
source it can be identifies with C~{\scriptsize VI}, although it seems
more likely that it is an effect of inaccurate modelling of the
contamination in XIS1.

\subsection{The 14 -- 40 keV PIN spectrum}

In addition to a soft excess and a broad Fe line, reflection from an
accretion disc should produce a Compton hump around 20--30~keV, as
seen in Fig.  \ref{2refmodel}. Data at these energies, provided by the
\suzaku HXD PIN detector, are therefore crucial for testing the models
derived from the XIS data. Specifically, the two-reflector model
predicts about 1.4 times more flux in the 14-40~keV range than the
smeared absorption model.

Mrk~335 is detected in the PIN up to about 40~keV and we stress that
this detection is robust against uncertainties in the background
model. However, since the background dominates the spectrum, the
accuracy of the background modelling is very important for our
results. As mentioned in section \ref{observations}, comparison of the
instrumental background model with night earth spectra shows that the
background model is overestimated by about 6 per cent. We therefore
correct for this before fitting the spectrum. We also include a model
for the CXB as described in the same section. The 14--40~keV flux from
the CXB model represents about 25 per cent of the source flux. The
cross normalization of the PIN with respect to the XIS detectors has
been reported to be 1.16 for the HXD nominal position and {\scriptsize
  V}1.2.2.3 data (Ishida et al. 2006), and we use this value in the
following analysis.

\begin{center}
\begin{figure}
\rotatebox{270}{\resizebox{!}{80mm}{\includegraphics{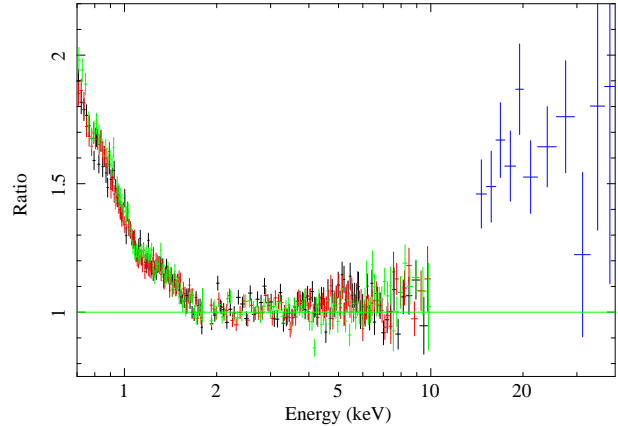}}}
\caption{The broad-band \suzaku spectrum of Mrk~335 shown as a ratio
  to the simple 2--10~keV power law + Gaussian model. Excess emission
  is clearly visible both at low and high energies. Data from the three
  FI XIS are shown in black, red and green. The PIN data are shown in
  blue.}\label{pin_ratio}
\end{figure}
\end{center}

\begin{center}
\begin{figure}
\rotatebox{270}{\resizebox{!}{80mm}{\includegraphics{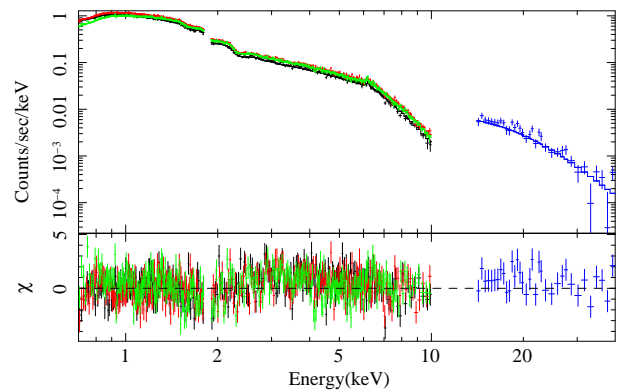}}}
\caption{ The broad-band \suzaku spectrum of Mrk~335.  The upper panel
  shows the data together with the two-reflector model shown in
  Fig. \ref{2refmodel}. The lower panel shows the residuals of the
  fit. Data from the three FI XIS are shown in black, red and
  green. The PIN data are shown in blue.}\label{spectrum}
\end{figure}
\end{center}

Fig. \ref{pin_ratio} shows the entire 0.7--40~keV spectrum as a ratio
to the simple phenomenological 2--10~keV model discussed in section
\ref{2to10}. In addition to the soft excess we see clear excess
emission in the PIN detector. In order to check if this emission can
be reproduced by the Compton hump of our 0.7--10~keV two-reflector
model we extrapolate this model to 40~keV. This gives
$\chi^2/\rm{d.o.f.}=5690/5410$ for the entire 0.7--40~keV
band. Fitting of the model with the PIN data included does not improve
the quality of the fit or change any of the parameter values. The fit
is shown in Fig. \ref{spectrum} and we see that the model slightly
underpredicts the PIN data. Given that the source is not very bright
in the hard X-rays, this slight discrepancy is likely to be due to
uncertainties in the background models (both non-X-ray background and
the CXB) and/or the cross-normalization of the PIN with respect to the
XIS detectors. If we instead extrapolate the blurred absorption model
to include the PIN data, we get $\chi^2/\rm{d.o.f.}=5734/5410$, i.e
$\Delta \chi^2 =43$ worse than for the two-reflector model. Fitting of
the absorption model with the PIN data included results in a slightly
higher contribution from the reflector, which improves the quality of
the fit by $\Delta \chi^2=4$.  The high-energy data hence favour the
two-reflector model, although both models are acceptable given the
uncertainties mentioned above.


\section{The time-averaged XMM-Newton spectrum}\label{xmmspec}

\begin{center}
\begin{figure}
\rotatebox{270}{\resizebox{!}{80mm}{\includegraphics{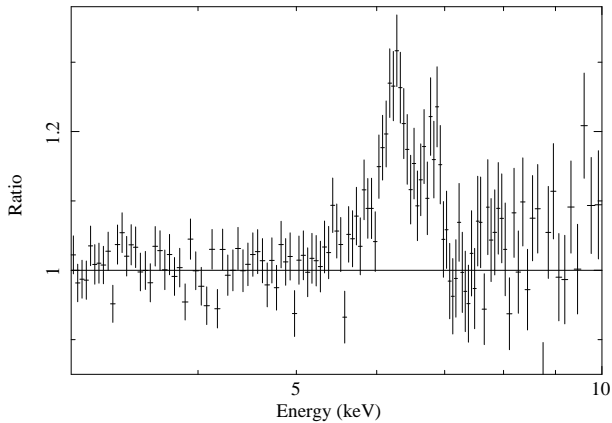}}}
\caption{The \xmmn pn spectrum shown as a ratio to a power law fitted
  over the 3--4.5 and 7.5--10 keV energy ranges. The Fe line profile
  is clearly double peaked, but apart from the blue peak it is very
  similar to the profile shown in Fig. \ref{fe_ratio} for the \suzaku
  data.}\label{xmmratio}
\end{figure}
\end{center}

Mrk~335 was observed for 133~ks by {\it XMM-Newton} about six months
before the \suzaku observation, and a comparison between these two
observations is likely to provide interesting information about the
source. Because \xmmn is more reliably calibrated at low energies, we
also expect that the \xmmn data will allow us to put better
constraints on the two-reflector model at low energies.

We start, however, by considering the 2--10~keV time-averaged
spectrum. The most notable difference compared to the \suzaku data in
this energy range is the very clearly double-peaked Fe line profile,
shown in Fig.  \ref{xmmratio}. Detailed fitting of the line profile
has been presented by O'Neill et al. (2007), who find that a
moderately broad relativistic line is present but that the two peaks
at 6.4 and 7~keV are most likely due to narrow lines, identified with
Fe K$\alpha$ and Fe {\scriptsize XXVI} Ly$\alpha$, respectively. The
two lines are suggested to originate in the molecular torus of AGN
unification models, and in highly ionized gas filling this torus,
respectively. Here we simply note that the Fe line profile can be
fitted with the same model as the \suzaku data if a narrow line is
added around 7~keV, in agreement with the findings of O'Neill et
al. (2007).

With the addition of the narrow line around 7~keV, the entire
2--10~keV \xmmn spectrum can be very well fitted with the models
presented for the \suzaku data. As an example, our model with
reflection arising from outside $40\ r_g$ in a disc with $\xi=30\
\rm{erg\ cm\ s^{-1}}$, gives $\chi^2/\rm{d.o.f.}=1324/1353$, with the
photon index and the relative normalization of the different
components changing very little from the \suzaku values. For the two
narrow lines we find energies of $6.41^{+0.02}_{-0.03}$~keV and
$7.05^{+0.03}_{-0.05}$~keV, with equivalent widths of $43\pm 13$~eV
and $32\pm 14$~eV, respectively. The 2--10~keV flux determined
from the model is $1.78\times 10^{-11}\ \rm{erg\ cm^{-2}\ s^{-1}}$,
compared to $1.43\times 10^{-11}\ \rm{erg\ cm^{-2}\ s^{-1}}$ for the
\suzaku data.  The flux in the narrow Fe~K$\alpha$ is consistent with
not having changed between the observations, with
$7.41^{+1.76}_{-1.53}\times 10^{-6}\ \rm{photons\ cm^{-2}\ s^{-1}}$
obtained for \suzaku and $6.58^{+1.44}_{-1.99}\times 10^{-6}\
\rm{photons\ cm^{-2}\ s^{-1}}$ for {\it XMM-Newton}.

At low energies, larger differences between the two data sets can
be seen. In terms of the two-reflector model, much of this difference
can be explained by increasing the ionization parameter of the inner
reflector from about $300\ \rm{erg\ cm\ s^{-1}}$ to about $1200\
\rm{erg\ cm\ s^{-1}}$ (no solution with the same ionization parameter
for the inner reflector in both data sets was found). However, in
order to get a good fit all the way down to 0.4~keV we also need to
include an edge at 0.51~keV with $\tau=0.33$ (as for XIS1, possibly
consistent with the 0.49~keV C~{\scriptsize VI} edge).  If we let all
the parameters of the model vary we find a best-fitting 0.4--10~keV
two-reflector (+ edge) model with $\chi^2/\rm{d.o.f.}=1749/1670$.
Apart from the higher ionization parameter for the inner reflector,
the only parameters in this model that differ significantly from the
\suzaku FI XIS fits are the photon index and the emissivity
index. These are both steeper; $\Gamma=2.28\pm 0.01$ and
$q=5.9^{+0.4}_{-0.6}$, respectively (we froze $q$ at 3 for the FI XIS
data). The reflection fraction and relative contribution of the two
reflectors in this model are very close to what we found for the
\suzaku data.

\section{Spectral variability}\label{variability}
\begin{center}
\begin{figure}
\rotatebox{270}{\resizebox{!}{80mm}{\includegraphics{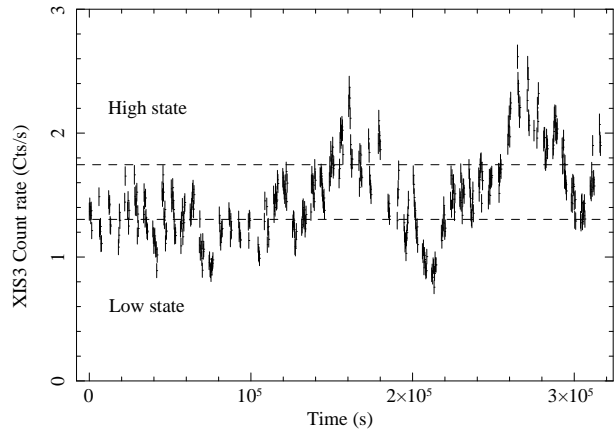}}}
\caption{The XIS3 lightcurve in the 0.5--12~keV band in 300~s
  bins. We also show the high and low flux states, which were defined
  so that they have roughly the same number of counts.}\label{lcurve}
\end{figure}
\end{center}

In Fig. \ref{lcurve}, we show the 0.5--12~keV background-subtracted
XIS3 light curve from the \suzaku observation of Mrk~335. The figure
also defines the high and low states which we will use to study the
spectral variability. The source varies by a factor of about 2.5
during the observation and the lightcurve notably contains two big
flares.

Below, we will investigate the spectral variability of Mrk~335 during
the \suzaku observation by considering hardness-ratios, the high and
low states, and rms spectra. Because of the high uncertainties in the
PIN background on short time-scales we limit this analysis to the XIS
data. At the end of the section we will compare our results with the
variability properties of the \xmmn data.

\subsection{Hardness-ratios}
\begin{center}
\begin{figure}
\rotatebox{270}{\resizebox{!}{80mm}{\includegraphics{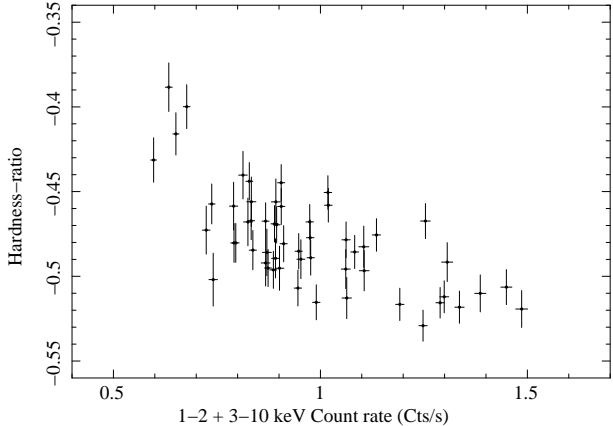}}}
\caption{The mean XIS hardness-ratio as a function of the mean count
  rate. The hardness-ratio is defined as $\rm{H-S/H+S}$, where H is
  the count rate in the 3--10~keV band and S is the count rate in the
  1--2~keV band. The source clearly hardens significantly at low count
  rates.  }\label{hratio}
\end{figure}
\end{center}
\begin{center}
\begin{figure}
\rotatebox{270}{\resizebox{!}{80mm}{\includegraphics{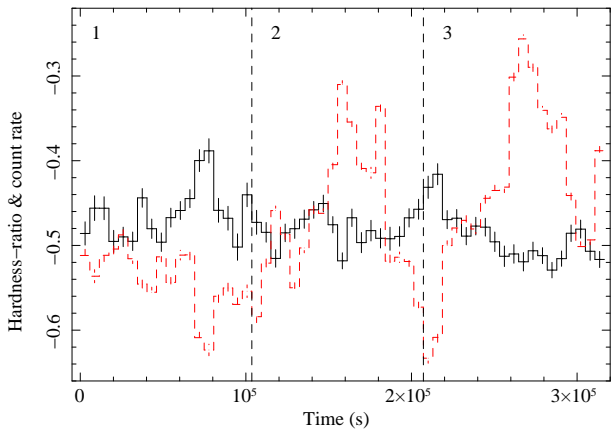}}}
\caption{The mean XIS hardness-ratio (black, solid line) as a function
  of time together with the re-scaled, mean XIS 0.5--12~keV lightcurve
  (red, dashed line). The source hardens significantly during the dips
  in the lightcurve but softens very little during the two big flares.
  The vertical dashed lines divide the observation into three time
  intervals of equal length. We will consider the rms spectra from
  these three intervals in section \ref{rms}.}\label{hr_time}
\end{figure}
\end{center}

As a first, model-independent, way of characterising the spectral
variability of the source, we calculate hardness-ratios,
$\rm{HR=H-S/H+S}$, where we take H to be the count rate in the
3--10~keV band and S to be the count rate in the 1--2~keV band. The
mean XIS hardness-ratio, calculated in orbital length bins of 5760~s,
is plotted as a function of the mean $\rm{H+S}$ count rate in
Fig. \ref{hratio}. At low count rates there is a clear
anti-correlation between the hardness-ratio and the count rate, but
this anti-correlation flattens out as the count rate increases. This
curious behaviour is even more evident in Fig. \ref{hr_time}, which
shows the mean XIS hardness-ratio as a function of time, together with
the re-scaled, mean XIS 0.5--12~keV lightcurve. We see that the source
hardens significantly during the dips in the lightcurve but that it
does not soften by a corresponding amount during the two big flares.

In Fig. \ref{hr_time}, we also divide the observation into three time
intervals of equal length. The hardness ratio in the middle part
varies less than it does in the first and the third parts (mainly
because this part of the lightcurve does not contain any large dip),
and in section \ref{rms} we will see how this leads to different
shapes of the rms spectra for the three intervals.

\subsection{High and Low states}\label{hl}
\begin{table*}
\begin{center}
\begin{tabular}{llrrr}
  \hline \\ [-5pt]
  Parameter & &\multicolumn{1}{c}{Low state} &
  \multicolumn{1}{c}{Intermediate state} &
  \multicolumn{1}{c}{High state} \\[5pt]
  \hline \hline
  \multicolumn{5}{c}{Power law + black-body + broad line + narrow line} \\
  \hline
  $\Gamma$ && $2.07\pm 0.02$ & $2.12\pm 0.02$  & $2.16\pm 0.02$\\
  PL Norm & $\times 10^{-3}$ &  $4.74^{+8.26}_{-0.10}$  &  $6.42\pm 0.10$  &
  $8.59\pm 0.20$\\
  BB Temp  &eV & $134^{+2}_{-5}$  &  135$\pm 4$  &  $142^{+3}_{-4}$\\
  BB Norm &$\times 10^{-4}$& $1.41^{+0.11}_{-0.10}$  & $1.76^{+0.10}_{-0.09}$   &
  $2.09^{+0.15}_{-0.14}$\\
  $\rm{E_{broad}}$ &keV & $6.54^{+0.14}_{-0.11}$ & $6.41^{+0.09}_{-0.08}$  &
  $6.42^{+0.13}_{-0.12}$ \\
  $\rm{EW_{broad}}$ &eV & $291^{+79}_{-74}$ & 294$^{+60}_{-57}$
  & 288$^{+85}_{-79}$\\
  $\rm{E_{narrow}}$ &keV & $6.36\pm 0.04$ & $6.39\pm 0.05$  &
  $6.38^{+0.12}_{-0.10}$ \\
  $\rm{Flux_{narrow}}$ &$\times 10^{-6} \rm{ph \ cm^{-2}\ s^{-1}}$
  & $5.43^{+3.87}_{-2.37}$ & $3.19^{+3.10}_{-2.12}$  & $5.04^{+4.65}_{-4.63}$ \\
 $\chi^2/\rm{d.o.f.}$ &  & 2818/2831 & 3870/3879 & 2693/2659  \\
  \hline
  \multicolumn{5}{c}{Power law + two reflectors + narrow line}\\
  \hline
  $\Gamma$  &&  2.14$\pm 0.02$ &  2.20$\pm 0.02$ & 2.26$\pm 0.02$\\
  PL/Tot flux$^{1}$ &(0.7-10~keV)  & 0.72  & 0.74  & 0.75\\
  $\rm{Ref_{inner}/Tot}$ flux$^{1}$ & (0.7-10~keV) & 0.22 & 0.19  &  0.18\\
  $\rm{Ref_{outer}/Tot}$ flux$^{1}$ & (0.7-10~keV) & 0.06  & 0.07  & 0.07\\
  $\rm{E_{narrow}}$ & keV & $6.37\pm0.03$ & $6.38\pm0.03$  &
  $6.37^{+0.07}_{-0.06}$\\
  $\rm{Flux_{narrow}}$ & $\times 10^{-6} \rm{ph \ cm^{-2}\ s^{-1}}$
  & $7.94^{+ 2.66}_{-2.44}$ & $6.22^{+ 2.30}_{-2.13}$ & $ 8.53^{+4.07}_{-4.00}$ \\
   $\chi^2/\rm{d.o.f.}$  & & 2800/2833 & 3826/3881 & 2665/2661 \\
  \hline
\end{tabular}
\caption{\label{table}\small{Results from fits to low-, intermediate-
    and high-state FI XIS spectra over the 0.7--10~keV band. The upper
    values are for a simple phenomenological model and the lower
    values are for the two-reflector model described in section
    \ref{softex}.  $ ^{1}$ Error bars are not given as it is currently
    not possible to calculate errors on fluxes of specific model
    components in} {\scriptsize XSPEC.}}
\end{center}
\end{table*}

In the previous section we have seen that there is an anti-correlation
between the spectral hardness and the flux of the source. In order to
investigate what spectral component(s) is(are) responsible for this
behaviour we now consider flux-selected spectra.  We select low-,
intermediate- and high-state FI XIS spectra as defined in
Fig. \ref{lcurve}. With these selection criteria, the high and low
states have about the same number of counts, corresponding to one
fourth of the total number of counts each. We fit both a
phenomenological model and the two-reflector model (described in
section \ref{softex}) to each of the three states over the 0.7--10~keV
energy band.

The phenomenological model consists of a power law, a black-body to
model the soft excess, and a broad and a narrow Gaussian line to model
the broad and narrow components of the Fe line. We let all parameters
vary apart from the widths of the Gaussian lines, which we freeze at
$\sigma_{\rm{broad}}=0.47\ \rm{keV}$ (as found for the time-averaged
fits in section \ref{2to10}) and $\sigma_{\rm{narrow}}=1\
\rm{eV}$. The results from the fits are presented in Table
\ref{table}. As expected, we see clear evidence of spectral pivoting,
with the photon index of the power law increasing from 2.07 in the low
state, to 2.12 in the intermediate state, and 2.16 in the high state.
(This linear increase of $\Gamma$ is simply an effect of how the flux
states were selected. When selecting a lower low state, we do indeed
see a much harder power-law, as expected from Figs. \ref{hratio} and
\ref{hr_time}.) The temperature of the black-body is found to be
roughly constant at 0.14~keV in all three states, while its
normalization increases with flux. Since the luminosity of a
black-body should scale with its temperature, this confirms that the
soft excess is not really a black-body.  For the broad component of
the Fe line, we find that the equivalent width stays nearly constant,
showing that this component varies together with the continuum. The
parameters of the narrow line are not very well constrained, but we
note that the flux of the line is consistent with being constant, as
would be expected if the line originates in distant material.

We next fit the three flux states with the two-reflector model from
section \ref{softex}. We freeze all parameters apart from $\Gamma$,
the energy of the narrow line, and the four normalizations (power law,
two reflectors and narrow line) at the time-averaged values found in
section \ref{softex}. The results from these fits are also presented
in Table \ref{table}. As for the phenomenological model, the power law
steepens with flux, with $\Gamma$ increasing from 2.14 in the low
state to 2.26 in the high state. Any attempts to fit the data with
$\Gamma$ frozen at the time-averaged value of 2.19 give fits which are
significantly worse than those presented in the Table.

As a measure of the relative contribution of the power law and the two
reflectors in the different flux states, Table~\ref{table} gives the
ratio of the flux in each of these components to the total 0.7--10~keV
flux.  We find that the outer reflector contributes by the same amount
(6-7 per cent) in all the flux states, while the inner reflector
contributes slightly more to the low state (22 per cent in the low
state compared to 18 per cent in the high state), indicating that it
is somewhat less variable.  Although this effect is quite small, we
note that a less variable inner reflector would be expected in a
picture where light bending is important close to the central black
hole (e.g. Miniutti \& Fabian 2004).

Since the spectral hardening is most prominent in the two deep dips of
the lightcurve we also extracted spectra in these intervals.
Specifically, we selected the 12~ks ($\sim$6~ks exposure) with the
lowest count rate in each of the dips. Fitting of these spectra are
consistent with the trends discussed above, although the short
exposure times make it difficult to unambiguously disentangle the
contribution from the different spectral components.

From the fits to both of the models presented above it seems clear
that the main driver behind the observed anti-correlation between the
hardness-ratio and the flux is a pivoting power law. Both the broad Fe
line and the soft excess appear to largely vary together with the
continuum, while the narrow component of the Fe line is consistent
with being constant.

\subsection{Rms spectra}\label{rms}
\begin{center}
\begin{figure}
\rotatebox{270}{\resizebox{!}{80mm}{\includegraphics{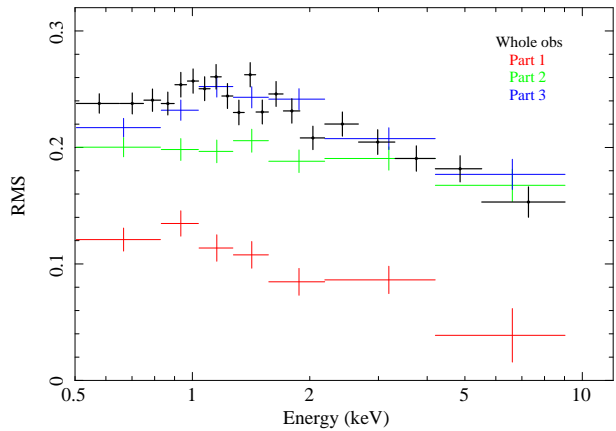}}}
\caption{Rms spectra of Mrk 335 on a 6~ks time-scale. The rms
  spectrum of the entire observation is shown in black, while rms
  spectra in red, green and blue were calculated for the first, second
  and third parts of the observation (as indicated in
  Fig. \ref{hr_time}). The middle part of the observation clearly
  gives a much flatter rms, consistent with the small variation in
  hardness-ratio seen in Fig. \ref{hr_time} during this time
  interval.}\label{suzrms}
\end{figure}
\end{center}

Another way of gaining insight into the variability properties of
Mrk~335 is the rms spectrum, which, for a given time-scale, shows the
fractional variability as a function of energy. The techniques for
calculating rms spectra are described in e.g. Edelson et al. (2002) and
Vaughan et al. (2003).

Fig. \ref{suzrms} shows rms spectra of Mrk~335, calculated on the
orbital time-scale of 5760~s. In order to investigate whether the rms
spectrum changes with time, we have calculated rms spectra from the
first, second and third parts of the observation (the three intervals
are defined in Fig. \ref{hr_time}) in addition to the rms spectrum of
the entire observation.  The errors from the Poisson noise were
calculated following Vaughan et al. (2003).

The rms spectrum of the entire observation (black crosses in
Fig. \ref{suzrms}) peaks around 1 keV and then decreases with energy,
a trend which is commonly observed in AGN.  This behaviour is often
explained either in terms of a pivoting power law, or in in a model
consisting of a constant reflection component and a power law that
varies in normalization but not in shape. It can also be explained in
an absorption model, assuming that the ionization parameter of the
absorber is driven by the continuum variability (Gierli\'{n}ski \&
Done 2006). In the case of Mrk~335, the fits to the different flux
states in the previous section suggest that the shape of its rms
spectrum is mainly due to a pivoting power law, but that a somewhat
less variable inner reflector may also contribute.

The rms spectra from the three parts of the observation show a change
in shape as well as in normalization, with the middle part being
significantly flatter than the first and the last parts, which both
have roughly the same shape as the rms of the whole observation.  The
flatter rms of the middle part is consistent with Fig. \ref{hr_time},
which shows that the hardness-ratio is less variable during this
period (the hardness only changes significantly during the large dips
in the lightcurve, none of which are present during the middle part).
The change of the shape of the rms spectrum is hence simply due to the
position and the length of the time interval that is being probed, and
does not imply that the source has entered a new mode of variability
or changed its spectrum (the time-averaged spectra of the three time
intervals are essentially identical). We also note that the flat shape
of the rms is difficult to explain within the absorption model, which,
assuming that the ionization state of the absorber responds to the
variable continuum, predicts an enhanced variability in the 0.8--2~keV
range (see also O'Neill et al. 2007).

In order to confirm that the two-reflector model is able to accurately
describe the variability, we also produced synthetic rms spectra for
the three intervals and the entire observation. This was done by
fitting the model to the individual 6~ks spectra and then using the
best-fitting models to extract fluxes in the different energy
bands. We find a good match between the observed and synthetic rms
spectra as long as we let the photon index and the normalizations of
the three components be free parameters in the fits, as expected from
the results of the previous section. Although this result is
encouraging, it should be noted that this simply means that our
two-reflector model provides a good fit to the entire energy range in
all the individual time intervals.

\subsection{Comparison with the XMM-Newton observation}
\begin{center}
\begin{figure}
\rotatebox{270}{\resizebox{!}{80mm}{\includegraphics{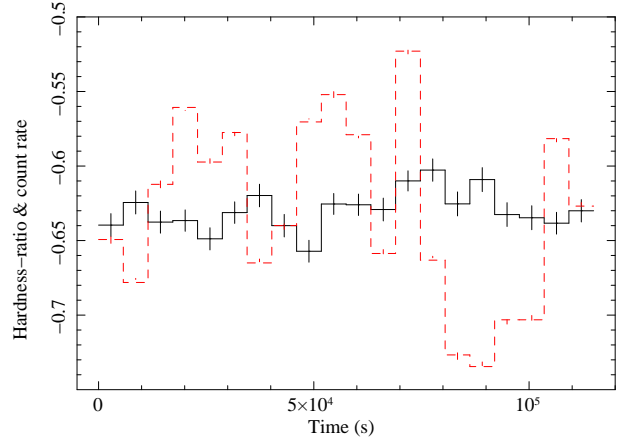}}}
\caption{Same as Fig. \ref{hr_time} but for the \xmmn
  observation. Note that both the lightcurve (red, dashed line) and
  the hardness-ratio (black, solid line) are much less variable than
  in the \suzaku observation.}\label{xmmhratio}
\end{figure}
\end{center}
\begin{center}
\begin{figure}
\rotatebox{270}{\resizebox{!}{80mm}{\includegraphics{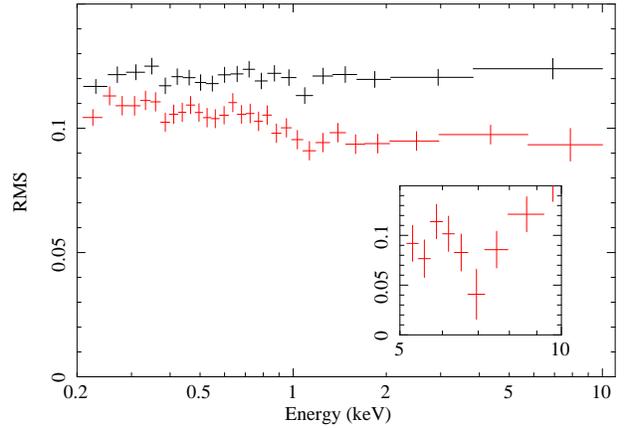}}}
\caption{Rms spectra of the entire \xmmn observation on a 1~ks (black)
  and a 6~ks (red) time-scale. The 1~ks rms is consistent with being
  constant at 0.12. The 6~ks rms decreases slightly with energy but it
  is much flatter than the rms of the whole \suzaku observation. The
  inset shows the 5--10~keV range of the 6~ks rms with a much finer
  energy binning. There is a clear dip in the variability at 7~keV,
  confirming that the 7~keV peak seen in the spectrum originates in
  distant material.}\label{xmmrms}
\end{figure}
\end{center}

In section \ref{xmmspec}, we have seen that the long \xmmn observation
of Mrk~335 catches the source in a very similar flux state as the
\suzaku observation does six months later.  The time-averaged spectra
of the two observations are very similar above 2~keV (apart from an
emission line at 7~keV only present in the \xmmn data), and we find
that similar two-reflector models fit both data sets also at lower
energies.  In this section we will extend the comparison between the
two observations by considering some of the variability properties of
the \xmmn data.

Fig. \ref{xmmhratio} shows the hardness-ratio together with the
re-scaled lightcurve of the \xmmn observation, calculated in the same
time bins as for the \suzaku observation (cf. Fig. \ref{hr_time}). We
note that both the count rate and the hardness-ratio are much less
variable than in the \suzaku observation, and that they do not show a
correlation.

From the almost constant hardness-ratio we would expect the rms
spectrum of the \xmmn observation to be fairly flat. This is exactly
what we see in Fig. \ref{xmmrms}, which shows the rms spectrum on both
a 1~ks and a 6~ks time-scale (the \suzaku rms spectra were calculated
on a 6~ks time-scale). The 1~ks rms spectrum is consistent with being
constant at 0.12, in excellent agreement with O'Neill et al. (2007).
The 6~ks rms decreases slightly with energy, but is still much flatter
than the rms spectrum of the entire \suzaku observation.  This
difference would naively suggest that the source has changed its
behaviour dramatically in the six months between the two observations.
However, we have seen that the \suzaku rms changes in shape during the
observation, and that the curved shape arises when the lightcurve
contains a large dip, in which the source hardens significantly.  It
therefore seems plausible that the rms spectrum is flat during the
\xmmn observation simply because the count rate never drops to a low
enough level. This interpretation is further supported by the fact
that the flux is slightly higher in the \xmmn observation and that it
is only a third as long as the \suzaku observation.

Fig. \ref{xmmrms} also shows the 5--10~keV range of the 6~ks rms with
a much finer energy binning, revealing a clear dip in the variability
at 7~keV. This supports our interpretation that the 7~keV peak seen in
the spectrum is a narrow line originating in distant material

\section{Discussion}\label{discussion}

Mrk~335 exhibits a strong soft excess below about 2~keV and does not
show any signs of complex, warm absorption at low energies. This makes
it an ideal target for studying the origin of the much debated soft
excess.

In this paper we have focused on explanations for the soft excess
which involve blurred reflection from an accretion disc. Such models
have been successful in fitting the spectra of previous observations
of Mrk~335 (e.g. Ballantyne et al. 2001; Crummy et al. 2006). In the
case of the recent \suzaku and \xmmn observations, we find that two
reflectors are required in order to simultaneously explain the strong,
smooth soft excess and the moderately broad Fe line.  Our best-fitting
model comprises a power law, an inner, heavily blurred, ionized
reflector (which produces most of the soft excess) and an outer,
almost neutral reflector (which is responsible for most of the Fe
emission).

The requirement for more than one reflector is a consequence of
several plausible disc/corona configurations.  A power-law source in
the form of a patchy corona, as well as a coronal geometry that
changes with time (e.g. due to magnetic processes), would for instance
give rise to different parts of the disc having different ionization
states.  However, about half of the reflected emission in our model
originates outside about 40~$r_g$, which seems implausible for a disc
with a patchy corona. A more centrally concentrated power-law source
could produce such a situation if it for example consists of two
sources at different heights, so that the inner and outer parts of the
disc are predominantly irradiated by the lower and upper sources,
respectively. Alternatively, the emission from a power-law source
located very close to the black hole could be beamed along the disc
(e.g. due to rotational motion about the black hole), which in
combination with flaring of the disc at large radii would lead to a
higher illumination of the outer disc.  We also note that models with
multiple reflectors have been successful in modelling the spectra of
several other Seyfert 1 galaxies, e.g. 1H 0707-495 (Fabian et
al. 2002) and PG 1211+143 (Crummy et al. 2006). It seems likely that
such models could be important in other sources that, like Mrk~335,
exhibit a strong soft excess in combination with a moderately broad Fe
line.

It is important to note that our two-reflector model for Mrk~335
relies on the fact that the inner reflector with $\xi\approx 300\
\rm{erg\ cm\ s^{-1}}$ produces a very weak Fe line.  This is because
the majority of the Fe in this ionization state is in the form of
L-shell ions, which in the {\scriptsize REFLION} model are assumed to
produce very little K$\alpha$ emission due to resonant Auger
destruction.  However, the amount of resonant Auger destruction
depends on the conditions in the accretion disc, and it is possible
that the fluorescent yield for several Fe L-shell ions actually remain
high (Liedahl 2005). It is not clear to what extent this would affect
our conclusions.

It should also be noted that the reflection model used for the
analysis in this paper assumes a constant density throughout the
disc. It has been pointed out by Done \& Nayakshin (2007) that the
soft excess is much weaker if a hydrostatic model for the disc is
used.  This is because the partially ionized material that produces
the soft excess can only exist in a thin layer in a hydrostatic
disc. However, it is not at all clear that accretion discs are in
hydrostatic equilibrium, as e.g. magnetic fields are likely to be
important. This therefore does not pose a serious problem for the
interpretation that the soft excess is due to reflection.

Our analysis of the spectral variability of Mrk~335 in terms of the
two-reflector model shows evidence for a pivoting power law and
suggests that both reflectors are varying together with the continuum
(although there is some evidence that the inner reflector is slightly
less variable). This behaviour is rather different from that observed
in the typical flux states of many other reflection-dominated sources,
such as MCG--6-30-15 (Miniutti et al. 2007), NGC 4051 (Ponti et
al. 2006) and 1H~0707 (Fabian et al. 2004). In these sources the
spectral variability can be described in terms of a power law that
varies in normalization (but not in slope) and a much less variable
reflection component. The spectral variability in Mrk~335 is also
unusual in that the biggest changes occur in the low flux states,
where the source hardens substantially. The reason for this unusual
behaviour is unclear and more data when the source is in a low state
is needed in order to properly study this variability.

As an alternative to the reflection model we have also explored the
possibility of the soft excess being produced as a result of smeared
absorption.  We have seen that the {\scriptsize SWIND1} absorption
model provides a very good fit to the spectrum of Mrk~335 as long as
we only consider the data up to 10~keV. However, the model
underpredicts the PIN data in the 14--40~keV range, suggesting that
more reflection is present in the spectrum. The observed flat rms
spectra of the \xmmn observation and part of the \suzaku observation
also disagree with the absorption model, as this model predicts
enhanced variability in the 0.8--2~keV range, assuming that the
ionization state of the absorber responds to the changing luminosity
of the source (Gierli\'{n}ski \& Done 2006).  As previously mentioned,
the {\scriptsize SWIND1} absorption model is also problematic in its
simplistic treatment of the velocity field, and it seems like a more
realistic model cannot produce the smooth shape of observed soft
excesses unless extreme velocities are incorporated (Schurch \& Done
2007). An absorption origin for the soft excess in Mrk~335 hence seems
very unlikely.

We have seen that the shape of the hard spectrum of Mrk~335 can be
very well explained with a blurred reflection model. However, it
should be pointed out that partial covering models have also been seen
to provide acceptable fits to the 3--10~keV energy band in e.g. the
{\it \xmmn} observations of the source (Longinotti et al. 2007;
O'Neill et al. 2007). We have not considered partial covering models
in our analysis but we note that the rapid variability of Mrk~335
implies that the emission region must be very small (less than about
one light hour across), and a partial coverer would hence have to be
even smaller and very dense. We also note that a broken power law (or
some other additional component) is required to explain the soft
excess within a partial covering model (Grupe et al. 2007), making
such a model less physically plausible.

Our comparison of the \xmmn and \suzaku observations has provided
interesting information about the origin of the narrow emission lines
seen in this object. The 6.4 keV line is clearly detected in both
observations and the flux in the line is consistent with being
constant, confirming that the line originates in distant material,
e.g. the molecular torus of AGN unification models. The 7~keV line
which is clearly seen in the \xmmn spectrum is, however, not detected
in the \suzaku data.  O'Neill et al. (2007) showed that the line
cannot be attributed to the blue horn of the broad Fe line, and our
own spectral and rms analysis support this
interpretation. Specifically, our rms analysis shows that the line
does not vary during the \xmmn observation, which places an inner
limit on the origin of the line at about a light day from the nucleus.
On the other hand, the fact that the line has varied in the six months
between the two observations places an outer limit on its origin at
around a light month. These constraints together suggest that the line
originates somewhere in the Broad Line Region of the source.

\section{Summary}\label{summary}

In this paper we have presented an analysis of a long \suzaku
observation of the NLS1 galaxy Mrk~335. We have also considered a
previous \xmmn observation of the source.

The \suzaku spectrum exhibits a broad Fe line and a strong soft
excess, and does not show any signs of warm absorption, in agreement
with previous observations. We find that a model consisting of a power
law and two reflectors provides the best fit to the entire 0.5--40~keV
time-averaged spectrum. In this model, a heavily blurred inner
reflector produces most of the soft excess, while an outer reflector
(outside $\sim 40\ r_g$) produces most of the Fe line emission.

The spectral variability of the \suzaku observation is characterised
by spectral hardening at low count rates. Fitting of different flux
states with the two-reflector + power-law model suggests that this
hardening is mainly due to pivoting of the power law. The two
reflectors appear to vary with the continuum, although there is some
evidence for the inner reflector being less variable.  The rms
spectrum of the entire observation has the curved shape commonly
observed in AGN, but the shape is significantly flatter when an
interval which does not contain any deep dips in the lightcurve is
considered. This suggests that the main driver behind the curved shape
of the rms is the spectral hardening which occurs in these dips.  The
physical reason for this hardening is currently unclear.

The time-averaged \xmmn spectrum can be fitted with a similar
two-reflector model as the \suzaku data and the two data sets have
similar fluxes. However, the \xmmn data do not show a correlation
between the spectral hardness and the count rate, and the rms spectrum
is flat. Since the spectral hardening of the \suzaku data mainly
occurs in the deep dips in the lightcurve, it seems plausible that the
lack of significant spectral variability in the \xmmn observation is
due to the fact that the count rate never drops to a low enough
level.

We have discussed the implications of our findings for the origin of
the soft excess in other Seyfert 1s, and we suggest that the
two-reflector model might be applicable in other sources that show a
soft excess which is stronger and/or smoother than expected from the
Fe line properties. We have also noted that the observed flat rms
spectra, as well as the high-energy data from the PIN detector,
disfavour an absorption origin for the soft excess in Mrk~335.

\section*{Acknowledgements}
JL thanks Corpus Christi College, the Isaac Newton Trust and STFC. ACF
thanks the Royal Society for support.


{}

\end{document}